\documentclass[%
 reprint,
superscriptaddress,
 amsmath,amssymb,
 aps,
floatfix,
]{revtex4-1}

\usepackage[english]{babel}
\usepackage[utf8x]{inputenc}
\usepackage[colorinlistoftodos]{todonotes}
\usepackage{enumerate}
\usepackage{graphicx}
\usepackage{dcolumn}
\usepackage{bm}

\usepackage{textcomp}
\usepackage[pagewise,columnwise]{lineno}

\newcommand{\avg}[1]{ \left\langle #1 \right\rangle }

\begin{document}

\title{Multiplicity dependence of the entropy and heat capacity for pp collisions at LHC energies}

\author{C. E. Munguía López}
\affiliation{Facultad de Ciencias F\'isico Matem\'aticas, Benem\'erita Universidad Aut\'onoma de Puebla, Apartado Postal 165, 72000 Puebla, Pue., M\'exico}

\author{D. Rosales Herrera}
\email{diana.rosales.herrera@cern.ch}
\affiliation{Facultad de Ciencias F\'isico Matem\'aticas, Benem\'erita Universidad Aut\'onoma de Puebla, Apartado Postal 165, 72000 Puebla, Pue., M\'exico}

\author{J. R. Alvarado García}
\email{j.ricardo.alvarado@cern.ch}
\affiliation{Facultad de Ciencias F\'isico Matem\'aticas, Benem\'erita Universidad Aut\'onoma de Puebla, Apartado Postal 165, 72000 Puebla, Pue., M\'exico}

\author{A. Fern\'andez T\'ellez}
\affiliation{Facultad de Ciencias F\'isico Matem\'aticas, Benem\'erita Universidad Aut\'onoma de Puebla, Apartado Postal 165, 72000 Puebla, Pue., M\'exico}

\author{J. E. Ram\'irez}
\email{jhony.eredi.ramirez.cancino@cern.ch}
\affiliation{Centro de Agroecología,
Instituto de Ciencias,
Benemérita Universidad Autónoma de Puebla, Apartado Postal 165, 72000 Puebla, Pue., M\'exico}

\begin{abstract}

We investigate the multiplicity dependence of the transverse momentum spectrum of the charged particle production in pp collisions at LHC energies.
To this end, we consider the experimental data sets classified with different multiplicity estimators, defined by the ALICE Collaboration, that are analyzed within the framework of nonextensive particle production. 
We compute the variance, kurtosis, Shannon entropy, and heat capacity of the $p_T$ spectrum to study the hardening process as a function of the multiplicity and temperature under the different event classifiers. 
We found that both the Shannon entropy and the heat capacity show different responses for the triggers at the forward-backward and midrapidity regions.
We emphasize that the selection of event biases may induce different responses in estimating theoretical and phenomenological observables that could lead to misleading conclusions.

\end{abstract}

\maketitle

\section{Introduction}
\label{sec:Intro}

The study of particle production in ultrarelativistic collisions provides insights into the physical phenomena occurring at very high energies, pressures, and temperatures \cite{BRAHMS:2004adc,PHOBOS:2004zne,STAR:2005gfr,PHENIX:2004vcz,ALICE:2022wpn}.
The transverse momentum spectrum of the produced particles is a specific experimental measure used to extract or infer information about the system created after the collision and its evolution 
\cite{VOGT2007221,Romatschke:2017ejr,ALICE:2013txf,ALICE:2015qqj,ALICE:2018vuu,Busza:2018rrf,ALICE:2022xip}. 
Fundamentally, 
the $p_T$ spectrum is a histogram built from the particles' yield observed after the freeze-out hadronic phase. 
It contains information about the particle production and its underlying physics
\cite{ALICE:2013wgn,ALICE:2016dei,CMS:2016fnw,ALICE:2016fzo,Nagle:2018nvi,Arleo:2022shs}.

Over time, various theoretical and phenomenological approaches have been developed to describe the $p_T$ spectrum based on different assumptions encompassing both perturbative and non-perturbative QCD. 
For instance, the factorization theorem, collinear factorization, parton distribution and fragmentation functions \cite{Collins:1987pm}, 
parton-hadron duality \cite{Veneziano:1968yb,Bjorken:1973gc}, and multiple parton interactions \cite{Diehl:2011yj}, among others.
In particular, we focus on those founded on statistical physics of large fluctuations to describe the experimental data 
\cite{Hagedorn:1983wk}.

One of the earliest approaches to describe the $p_T$ spectrum data of the charged particle production was a $p_T$-exponential, which resembles the Boltzmann distribution
\cite{Hagedorn:1965st,Hagedorn:1967tlw,Frautschi:1971ij1}. 
In this approach, the inverse of the decay constant is identified as the effective temperature of the medium formed after the collision \cite{Braun-Munzinger:1994ewq,Garcia:2022ozz}.
Even though the $p_T$-exponential captures the experimental behavior of the $p_T$ spectrum data for pp collisions at very low center of mass energies or in the low $p_T$ limit, it fails to describe the hard part of the $p_T$ spectrum, which substantially deviates from the data as the center of mass energy grows.
The latter inspired several authors to introduce quasi-power law functions to describe the $p_T$ spectrum~\cite{Hagedorn:1983wk,Michael:1976pz,Bhattacharyya:2017cdk}. 
In particular, the Hagedorn's proposal stands out above all of them because it is a QCD-inspired function that appropriately describes the asymptotic limits of the $p_T$ spectrum observed experimentally \cite{Hagedorn:1983wk}. 
The Hagedorn function behaves as a $p_T$-exponential function at low $p_T$ and incorporates the power law trend at high $p_T$ values.
Later, the high energy particle physics community adopted Tsallis $q$-exponential functions as generalizations of the $p_T$-exponential distribution \cite{Wilk:1999dr,Biro:2016myk,Bhattacharyya:2017hdc,Saraswat:2017kpg,Yang:2019rje,Herrera:2024tyq},
having a good performance in fitting the experimental $p_T$ spectrum data, leading to infer information about the systems created in ultrarelativistic particle collisions \cite{Mishra:2021hnr}. 
However, it is phenomenologically equivalent to the Hagedorn function
\cite{Wong:2015mba,Rybczynski:2020kvl}.

Another approach comes from the color string models, which are frameworks frequently used to study the particle production through the fragmentation of color strings stretched between the colliding partons \cite{andersson1998lund}. 
One way to describe the transverse momentum spectrum of the produced particles is by incorporating the Schwinger mechanism \cite{Schwinger:1962tp} together with string tension fluctuations. 
For instance, the thermal distribution can be obtained by 
assuming a Gaussian description for the string tension fluctuations \cite{Bialas:1999zg}. Nevertheless, heavy-tailed distributions should be considered in order to obtain a $p_T$ spectrum with adequate asymptotic limits: thermal distribution and a power-law tail at low and high $p_T$ values, respectively \cite{Pajares:2022uts}. This 
consideration leads to a nonextensive description of the particle production and the medium formed in the collision system, marking the departure from the thermal description \cite{Garcia:2022eqg,Herrera:2024zjy,Herrera:2024tyq}.

One way to study the processes involved in the charged particle production is by analyzing the $p_T$ spectra under different event classifications. This can be done by computing the Shannon entropy for the normalized 
$p_T$ spectrum. For instance, in Ref. \cite{Herrera:2024zjy}, the authors computed the Shannon entropy for the 
$p_T$ spectrum of the produced charged particles in minimum bias pp collisions as a function of the center of mass energy ($\sqrt{s}$). It was found that the Shannon entropy increases as a function of $\sqrt{s}$. The latter is expected not only because of the increase in the production of soft particles (which increases the temperature of the medium) but also because of the 
hardening of the 
$p_T$ spectrum 
(greater production of high $p_T$ particles).

In this work, we aim to analyze the multiplicity dependence of the Shannon entropy and heat capacity of the $p_T$ spectrum for the charged particles produced in pp collisions at $\sqrt{s}$ = 5.02, 7, and 13 TeV, considering the ALICE V0M and SPD multiplicity classifiers \cite{ALICE:2018pal,ALICE:2019dfi}. 
To this end, we adopt the nonextensive particle production approach, 
considering the Schwinger mechanism with heavy-tailed string tension fluctuations to describe the $p_T$ spectrum data. 
One consequence of this approach is that the particle production departs from the thermal equilibrium \cite{Herrera:2024tyq}.

We consider the normalized $p_T$ spectrum as a probability density function to compute its moments and the Shannon entropy. 
The latter is useful to study the dependence of the $p_T$ spectrum on the multiplicity classifiers.
Variations are expected since the V0M and SPD classifications are defined from the detectors' signals allocated at two complementary regions, forward-backward and midrapidity, respectively.

The forward-backward region is sensitive to initial collision scatters and long-range correlations, 
providing information about the longitudinal fragmentation process and the structure of the colliding particles \cite{CMS:2010ifv}. 
It is important to note that this region is used to determine the centrality of events in heavy-ion collisions using the Glauber model \cite{Miller:2007ri}. 
Additionally, the ALICE Collaboration uses the V0M multiplicity estimator as a strategy that may minimize the possible auto-correlations induced by using the midrapidity estimator \cite{ALICE:2013tla,ALICE:2019dfi}. 
For instance, the transverse jet fragmentation and event multiplicity may be correlated because they are measured within the same rapidity interval \cite{Ortiz:2016kpz}.
In contrast, the multiplicity estimator defined in the midrapidity region 
mainly captures the information of particles created in
earlier stages of the medium formed in the collision, which is useful for studying its properties \cite{CMS:2016fnw,Nagle:2018nvi}.

We emphasize that this manuscript is focused on studying the evolution of the 
$p_T$ spectrum across multiplicity classes, highlighting the 
entropy and heat capacity responses in each event classifier.

The rest of the paper is organized as follows. 
In Sec.~\ref{sec:TMD}, we discuss the derivation of the $p_T$ spectrum from a nonextensive particle production perspective. 
In Sec.~\ref{sec:Data}, we comment on the methodology and results of the fits to the experimental data for different energies and multiplicity classes. 
We also discuss the correlations between model parameters and their temperature dependence.
In Sec.~\ref{Sec:nmoments} we compute the $n$-th moments of the $p_T$ spectrum.
In Sec.~\ref{sec.HC}, we show the results of the Shannon entropy and the heat capacity of the analyzed 
$p_T$ spectrum data. 
A discussion on the physical interpretation of our finding is included in Sec.~\ref{sec.Disc}. 
Finally, we wrap up this work with our concluding remarks and perspectives in Sec.~\ref{sec.Conc}.

\section{Nonextensive origin of the transverse momentum spectrum}
\label{sec:TMD}

In string models, the particle production in ultrarelativistic collisions is basically based on the fragmentation of the color strings stretched between the partons in the colliding projectiles. In this way, the transverse momentum of the produced particles is modeled by the Schwinger mechanism \cite{Schwinger:1962tp}, given by
\begin{equation}
\frac{d N}{d p_T^2}\propto e^{-\pi p_T^2/x^2},
\label{eq:Schwinger}
\end{equation}
where $dN/dp_T^2$ denotes the 
$p_T$ spectrum.
Equation~\eqref{eq:Schwinger} can be interpreted as the probability of producing a particle with transverse momentum $p_T$ coming from the fragmentation of a color string with tension $x^2$. 
In a general scenario, the string tension may fluctuate \cite{Bialas:1999zg,Pajares:2022uts,Garcia:2022eqg}. 
Then, the $p_T$ spectrum should be computed as the following marginal probability 
\begin{equation}
    \frac{d N}{d p_T^2}\propto \int_0^\infty e^{-\pi p_T^2 / x^2}P(x)dx,
\label{eq:convolution}
\end{equation}
where $P(x)$ is the probability distribution function for the string tension fluctuations.

It has been shown that the Schwinger mechanism convoluted with a heavy-tailed distribution for the string tension fluctuations gives an adequate description for particle production in pp collisions \cite{Pajares:2022uts,Garcia:2022eqg,Herrera:2024zjy}.
The main implication of this approach is that the high $p_T$ particle production requires a nonextensive description of the initial state, and then, the medium formed is no longer thermal \cite{Herrera:2024tyq}. 
These phenomena can be well-modeled by using the $q$-exponential function, given by
\cite{abe}
\begin{equation}
    e_q(x)=[1+(1-q)x]^{1/(1-q)}.
\end{equation}
In particular, if $P(x)$ is a $q$-Gaussian distribution \cite{budini2}
\begin{equation}
    P(x)= 
    \mathcal{N}_q 
    e_q\left( - x^2 / 2\sigma^2 \right),
    \label{eq:qGauss}
\end{equation}
where $\mathcal{N}_q $ is the normalization constant, Eq.~\eqref{eq:convolution} becomes
\begin{equation}
    \frac{dN}{dp_T^2} 
    \propto U( a, 1/2, z_0 p_T^2  ),
\label{eq:TMDU}
\end{equation}
with $a=\frac{1}{q-1}-\frac{1}{2}$ and $z_0=\pi(q-1)/2\sigma^2$ \cite{Pajares:2022uts}.
In \eqref{eq:TMDU}, $U$ denotes the confluent hypergeometric function (also
known as the Tricomi's function), defined as
\begin{equation}
    U(a, b, z)=  \frac{1}{\Gamma(a)}\int_0^\infty e^{-zt} t^{a-1}(1+t)^{b-a-1} dt.
    \label{eq:U}
\end{equation}
Remarkably, the Tricomi's function~\eqref{eq:TMDU} has the appropriate asymptotic behaviors observed experimentally. At low $p_T$ values, Eq.~\eqref{eq:TMDU} recovers the thermal distribution $dN/dp_T^2 \propto e^{-p_T/T_U}$.  
We adopt the use of the inverse of the constant decay as the soft scale of the $p_T$ spectrum, usually interpreted as an effective temperature, given by
\begin{equation}
T_U= \frac{B(a , 1/2)}{\sqrt{4\pi z_0}},
\label{eq:TU}
\end{equation}
where $B$ is the beta function. $T_{U}$  can be understood as the temperature linked to the $p_T$ spectrum, computed over the ensemble of collision events occurring under the same biases \cite{Garcia:2022eqg}.
We must emphasize that the thermal distribution is also obtained from the Schwinger mechanism by assuming Gaussian fluctuations of the string tension \cite{Bialas:1999zg}, which is also recovered by taking the limit $q\to 1$ in the $q$-Gaussian distribution \eqref{eq:qGauss}.
Another limit of interest occurs when $p_T$ takes high values. In this limit, Eq.~\eqref{eq:TMDU} becomes the following power law $dN/dp_T^2 \propto (p_T^2)^{\frac{1}{2}-\frac{1}{q-1}}$.
Note that $q$ modulates the exponent of the 
$p_T$ spectrum tail. In this way, larger $q$ values mean bigger production of high $p_T$ particles and the hardening of the $p_T$ spectrum. In particular, for the production of charged particles in minimum bias pp collision, it was found that $q$ consistently grows with the center of mass energy \cite{Herrera:2024zjy}.

\begin{figure*}
    \centering
    \includegraphics{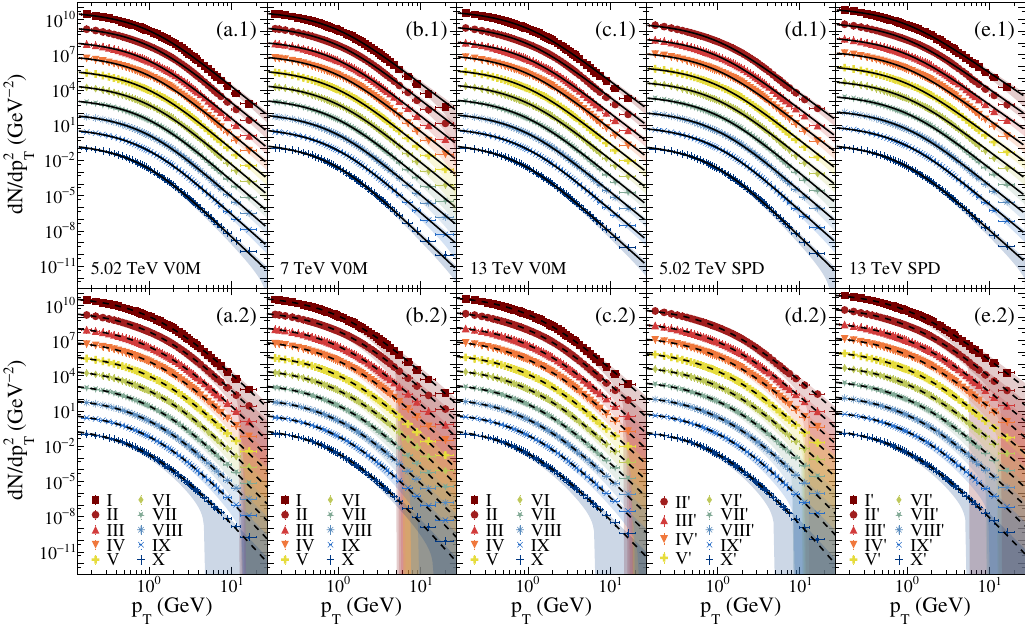}
        \caption{Experimental data of $p_T$ spectrum (markers) of the charged particle production in pp collisions at different center of mass energies under the V0M and SPD classifiers defined by the ALICE Collaboration: (a.1)-(a.2) $\sqrt{s}=$ 5.02 TeV (V0M), (b.1)-(b.2) $\sqrt{s}=$ 7 TeV (V0M), (c.1)-(c.2) $\sqrt{s}=$ 13 TeV (V0M), (d.1)-(d.2) $\sqrt{s}=$ 5.02 TeV (SPD), and (e.1)-(e.2) $\sqrt{s}=$ 13 TeV (SPD). 
        See Table \ref{tab:dsets} for detailed information on the kinematic cuts of the data sets. The solid lines correspond to the fits to data performed 
        via Eq.~\eqref{eq:TMDU} (top panels) and the dashed lines correspond to the fits using Eq.~\eqref{eq:qexp} (bottom panels). In all cases, the shaded region corresponds to the 2-$\sigma$ uncertainty propagation.} 
    \label{fig:fits}
\end{figure*}

As we commented in the introduction section, the Hagedorn function represents an alternative approach for describing the $p_T$ spectrum data of charged particle production \cite{Hagedorn:1965st,Hagedorn:1964zz,Hagedorn:1983wk}, and its phenomenology is equivalent to the Tsallis $q$-exponential distribution~\cite{Wong:2015mba,Saraswat:2017kpg} which is given by~\cite{Bialas:2015pla,Deb:2019yjo,Biro:2020kve}

\begin{equation}
\frac{dN}{dp_T^2} \propto
e_{q_e} \left( - p_T/T_e \right),
\label{eq:qexp}
\end{equation}
where $q_e$ and $T_e$ are parameters related to the nonextensivity and the soft scale of the system's final state, respectively. Here, we introduce the subscript $e$ to distinguish these parameters from other instances of $q$ used in the context of $q$-Gaussian string tension fluctuations. 
The mapping between the Hagedorn function $dN/dp_T^2 \sim (1 + p_T/p_0)^{-m}$ and Eq.~\eqref{eq:qexp} is the following: $p_0 = m T_e$, and $m= 1/(q_e-1)$.
At low $p_T$, Eq.~\eqref{eq:qexp} behaves as a thermal distribution, with $T_e$ being the inverse of the decay constant. At high $p_T$, it follows a power-law $\sim p_T^{1/(1-q_e)}$. 
Similar to the confluent hypergeometric function $U$, the Tsallis distribution also matches the asymptotic behaviors observed experimentally. 
 
Interestingly, Wilk and W\l{}odarczyk expressed the Tsallis distribution as the convolution of the thermal distribution with nonlocal temperature fluctuations ($\mathcal{T}$)~\cite{Wilk:1999dr}, which is useful to derive the string tension fluctuations $P_e(x)$ associated with Eq.~\eqref{eq:qexp}. 
Since the thermal distribution is obtained by the convolution of the Schwinger mechanism with Gaussian string tension fluctuations, $P_e(x)$ can be derived by convoluting the Gaussian distribution with $\mathcal{T}$ \cite{Herrera:2024tyq}. 
Therefore
\begin{equation}
e_{q_e} \left( - p_T/T_e \right)
= \int_0^\infty e^{-\pi p_T^2 / x^2} P_\text{e}(x) dx,
\label{eq:qStension}
\end{equation}
where $P_\text{e}(x)$ is given by
\begin{equation}
P_\text{e}(x) 
= \frac{ 1}{ \pi T_e}
\left[\frac{ 4 a_e^2 \pi T_e^2}{q_e^2 x^2} \right]^{a_e} U\left( a_e, \frac{1}{2} , \frac{4a_e^2\pi T_e^2}{q_e^2 x^2}  \right),
\label{eq:PHag}
\end{equation}
with $a_e = q_e/2(q_e-1)$.
For small $x$, $P_e(x)$ approximates to $\exp[- x^2 (2q_e^2 - q_e) / 2\pi T_e^2 ]$. In contrast, for large $x$, $P_e(x) \sim x^{-2a_e}$. 
In consequence, $P_e(x)$ is a heavy-tailed distribution~\cite{Herrera:2024tyq}.

We remark that the $U$ function \eqref{eq:TMDU} is derived by considering that the string tension fluctuations obey a $q$-Gaussian distribution, which is a nonextensive description of the initial state.
In contrast, the Tsallis $q$-exponential distribution is, a priori, proposed to describe the $p_T$ spectrum based on the nonthermal arguments of the final state.
Interestingly, as we discussed, both approaches can be derived from heavy-tailed distributions for the string tension fluctuations.
In this context, the emergence of a heavy-tailed $p_T$ spectrum implies the existence of phenomena out of thermal equilibrium, a necessary requirement for the production of high $p_T$ particles \cite{Garcia:2022eqg,Herrera:2024zjy,Herrera:2024tyq}.
This connection suggests that nonextensive statistical physics provides a framework for describing a wide range of effects in high-energy collisions, whether through the fragmentation of color strings or temperature fluctuations. 

In Sec.~\ref{sec:Data}, we analyze the characteristics of the $q$-exponential distribution and the $U$ function in describing the $p_T$ spectra, categorized by V0M and SPD multiplicity classes in pp collisions, and provide a correlation between the model parameters of both approaches.

\section{Data analisys}
\label{sec:Data}

We analyze the experimental 
$p_T$ spectrum data of the charged particle production in pp collisions reported by the ALICE Collaboration at midrapidity using the V0M and SPD multiplicity classes. 

\begin{table}[ht]    
 \centering
    \caption{Description of the experimental 
    $p_T$ spectrum data of the charged particle production in pp collisions analyzed in this manuscript. The data is reported by the ALICE Collaboration at midrapidity, classified according to the V0M and SPD definitions at different center of mass energies.}    
    \centering    
    \begin{ruledtabular}
    \begin{tabular}{cccc} 
    Data set & Rapidity cut & $p_T$ range& Ref.\\ \hline
    V0M at 5.02 GeV & $|\eta|<0.8$ & 0.15-20 GeV & \cite{ALICE:2019dfi} \\ 
    SPD at 5.02 GeV& $|\eta|<0.8$ & 0.15-20 GeV & \cite{ALICE:2019dfi} \\
    V0M at 7 TeV & $|\eta|<0.5$ & 0.16-40 GeV & \cite{ALICE:2018pal} \\
    V0M at 13 TeV& $|\eta|<0.8$ & 0.15-20 GeV & \cite{ALICE:2019dfi} \\
    SPD at 13 TeV& $|\eta|<0.8$ & 0.15-20 GeV & \cite{ALICE:2019dfi} 
    \end{tabular} \label{tab:dsets}     
    \end{ruledtabular}
\end{table}

In Table~\ref{tab:dsets}, we provide detailed information on the kinematic cuts of each data set.
We fit Eqs.~\eqref{eq:TMDU} and~\eqref{eq:qexp} to the entire $p_T$ range reported experimentally using the ROOT 6 software. 
In Fig.~\ref{fig:fits}, we show the fits to the $p_T$ spectrum data. 
In all cases, the ratio $\chi^2/$NDF takes values around 1, meaning that both approaches provide a good description of the experimental data for all the $p_T$ range reported. 
However, the error propagation via the model parameters is much better controlled by Eq.~\eqref{eq:TMDU}, as seen in Fig.~\ref{fig:fits}. 
In all cases, we assumed that the fitting parameters are independent.

\begin{figure}
    \centering
    \includegraphics{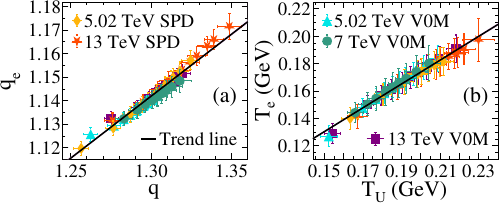}
    \caption{Correlations between the model parameters (a) $q_e$ as a function of $q$ and (b) $T_e$ as a function of $T_U$ for all the analyzed cases. The solid lines correspond to the linear trends.}
    \label{fig:parcor}
\end{figure}

We found that the nonextensivity parameters are linearly correlated by $q_e = \lambda_q q + b_q$ with $\lambda_q=0.53(1)$ and $b_q=0.46(1)$ (see Fig.~\ref{fig:parcor} (a)), which is consistent with previous parametrizations \cite{Herrera:2024tyq}. 
On the other hand, the model temperatures follow $T_e = \lambda_T T_U$ with $\lambda_T = 0.868(1)$ (see Fig.~\ref{fig:parcor} (b)).

\begin{figure*}[ht!]
    \centering
    \includegraphics{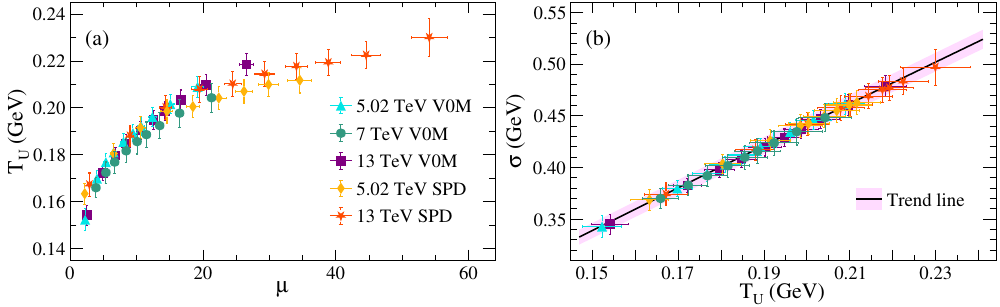}
    \caption{(a) Temperature as a function of the average multiplicity and (b) $\sigma$ as a function of temperature for all the analyzed data sets. The solid line corresponds to Eq.~\eqref{eq:sT}. The shaded region corresponds to the uncertainty propagation.}
    \label{fig:muT}
\end{figure*}

\begin{figure*}[ht!]
    \centering
\includegraphics{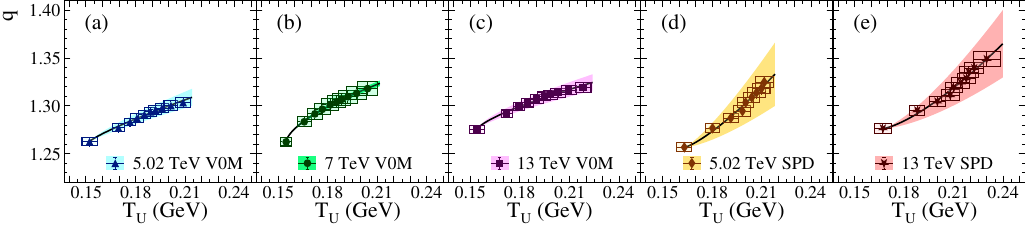}    
    \caption{Dependence of the $q$ on the temperature for: (a) $\sqrt{s}=$ 5.02 TeV (V0M), (b) $\sqrt{s}=$ 7 TeV (V0M), (c) $\sqrt{s}=$ 13 TeV (V0M), (d) $\sqrt{s}=$ 5.02 TeV (SPD), and (e) $\sqrt{s}=$ 13 TeV (SPD). The markers correspond to the fitted value of each data set. The solid lines correspond to the parametrization in Eq.~\eqref{eq:qT}. The shaded regions correspond to the uncertainty propagation.}
    \label{fig:qT}
\end{figure*}

Additionally, the temperature $T_U$ is a monotonically increasing function of the average multiplicity density ($\mu$), as seen in Fig.~\ref{fig:muT} (a). 
Therefore, it is possible to use $T_U$ as the main natural variable.
In all cases, $\sigma$ and $q$ increase with the multiplicity, and then, with $T_U$ (see Figs.~\ref{fig:muT} (b) and~\ref{fig:qT}). 

Moreover, $\sigma$ follows a global trend given by
\begin{equation}
    \sigma(T_U) = a_\sigma T_U^{c_\sigma},
    \label{eq:sT}
\end{equation}
where $c_\sigma = 0.868(1)$ and $a_\sigma = 1.94(2) \text{ GeV}^{1-c_\sigma}$. 
Relation~\eqref{eq:sT} is valid for all the analyzed data, as shown in Fig.~\ref{fig:muT} (b).

Figure~\ref{fig:qT} shows the dependence of $q$ as a function on the temperature for different center of mass energies and multiplicity classifiers.
We also propose the following 
parametrization
\begin{equation}
    q(T_U) = q_0 + a_q (T_U - T_0)^{c_q},
\label{eq:qT}
\end{equation}
where $q_0$ and $T_0$ are the minimal values of $q$ and $T_U$, corresponding to the lowest multiplicity class of each estimator. 
The obtained values of $a_q$ and $c_q$ are reported in Table~\ref{tab:t1}.

\begin{table}[ht]
    \centering
    \caption{Values obtained of the relation $q$ vs $T_U$ in Eq.~\eqref{eq:qT}.}    
    \begin{ruledtabular}
    \begin{tabular}{c c c c} 
       $\sqrt{s}$ (TeV)  & Classifier & $a_q$  $(\text{GeV}^{-c_\sigma})$ & $c_q$ \\ \hline 
5.02	&	V0M	&	0.47	$\pm$	0.07	&	0.83	$\pm$	0.05	\\ 
7	&	V0M	&	0.37	$\pm$	0.01	&	0.63	$\pm$	0.01	\\ 
13	&	V0M	&	0.33	$\pm$	0.04	&	0.72	$\pm$	0.04	\\ 
5.02	&	SPD	&	3.24	$\pm$	1.04	&	1.29	$\pm$	0.10	\\ 
13	&	SPD	&	3.38	$\pm$	1.02	&	1.39	$\pm$	0.10	
    \end{tabular}
    \label{tab:t1}
    \end{ruledtabular}
\end{table}

Equation~\eqref{eq:qT} captures some dissimilarities between the multiplicity classifiers through the dependence of $q$ on the temperature. 
For instance, for the V0M classes, $q$ exhibits a concave-down increasing behavior with $T_U$ (see Figs~\ref{fig:qT} (a)-(c)), which leads to an apparent saturation of $q$ as the multiplicity increases.
This means that there is no substantial increment in the production of high $p_T$ particles in the midrapidity region when the V0 activity grows. 
In contrast, for the SPD classifier, $q$ exhibits a concave-up increasing behavior with $T_U$, as shown in Figs~\ref{fig:qT} (d)-(e). 
This means that an increase in the multiplicity brings an increment in the production of high $p_T$ particles.
The last observation is consistent with the $p_T$ auto-correlation biases implicitly present in the SPD classifier, because the multiplicity classification is determined in the same region where the $p_T$ spectrum is measured \cite{ALICE:2018pal}, as we commented before.

\section{Moments and statistics of the transverse momentum spectrum}
\label{Sec:nmoments}

\begin{figure*}[ht!]
    \centering
\includegraphics{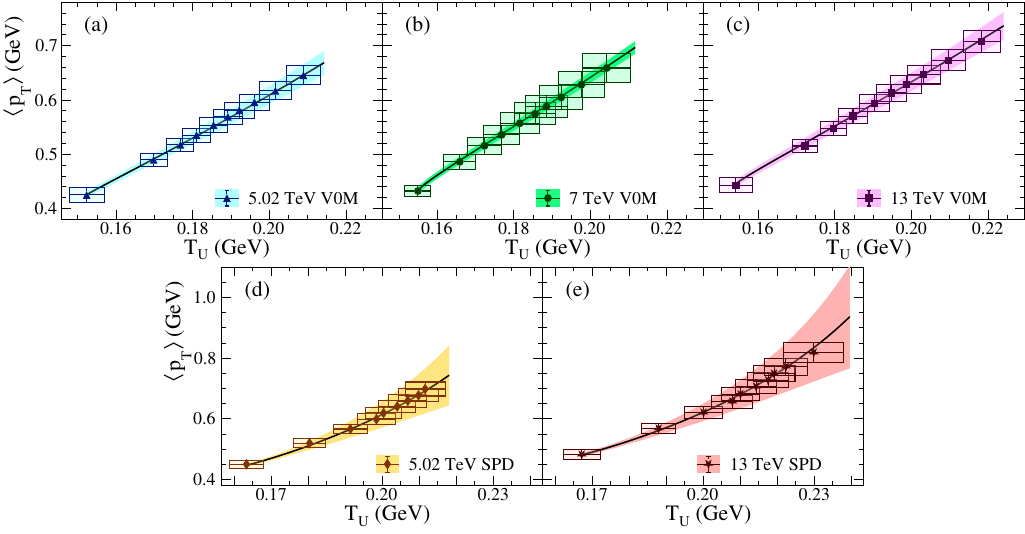}\caption{Temperature dependence of the mean $p_T$ (Eq.~\eqref{eq:avgpTe}) of the charged particles produced in pp collisions at (a)  
    $\sqrt{s}=$ 5.02 TeV (V0M classes), (b) $\sqrt{s}=$ 7 TeV (V0M classes), (c) $\sqrt{s}=$ 13 TeV (V0M classes), (d) $\sqrt{s}=$ 5.02 TeV (SPD classes), and (e) $\sqrt{s}=$ 13 TeV (SPD Classes). The markers correspond to the computed values using the values of $q$ and $\sigma$ for each data set. The solid lines correspond to the parametrization considering the temperature dependence of $q$ and $\sigma$ via Eqs.~\eqref{eq:sT} and~\eqref{eq:qT}). The shaded regions correspond to the uncertainty propagation.}
    \label{fig:meanpT}
\end{figure*}

The statistical moments of the $p_T$ spectrum provide characteristics on the central tendency, dispersion, and contribution of the tail to the whole spectrum, obtaining insights into the evolution of the $p_T$ spectrum as a function of the average multiplicity. 
 
We compute the $n$-th moments of the normalized $p_T$ spectrum as follows
\begin{equation}
\langle \mathcal{P}_T^n \rangle
= \frac{ \int  p_T^n  \frac{dN}{dp_T^2} dp_T}{\int \frac{dN}{dp_T^2} dp_T} .
\label{eq:ptnA}
\end{equation}
By computing the integral from 0 to infinity, Eq.~\eqref{eq:ptnA} can be expressed in the following closed form 
\begin{equation}
\begin{split}
   \avg{\mathcal{P}_T^n}   
= & \pi^{\frac{n-1}{2}}2^n \Gamma \left( \frac{n+1}{2} \right) \left( \frac{2-q}{q-1} \right) 
  \\ & \times 
  \frac{ B \left( \frac{n+2}{2}, \frac{1}{q-1}- \frac{n+2}{2} \right) }{ B \left( \frac{1}{q-1} -\frac{1}{2}, \frac{1}{2}\right)^n } T_U^n, 
\label{eq:ptn}  
  \end{split}
\end{equation}
which is well-defined for $q<({4+n})/({2+n})$ \cite{Herrera:2024zjy}. 

Additionally, the computation of the moments reported by the experiments requires multiplying the factor $2\pi p_T$ from phase space to the $p_T$ spectrum \cite{Hagedorn:1983wk}. 
Thus, we found 
\begin{equation}
      \avg{p_T^n}= \avg{\mathcal{P}_T^{n+1}}/\avg{\mathcal{P}_T}.
\label{eq:momentslab2}  
\end{equation}

It is important to 
emphasize that Eq.~\eqref{eq:ptnA} can also be computed over the $p_T$ range reported experimentally, which is a kinematic cut defined by the setup where the experiments have an optimal resolution. 
In those cases, our estimations match the data reported by the experiments, as discussed in Ref.~\cite{Herrera:2024zjy}.

\subsection{Average transverse momentum}

\begin{figure}
    \centering
 \includegraphics{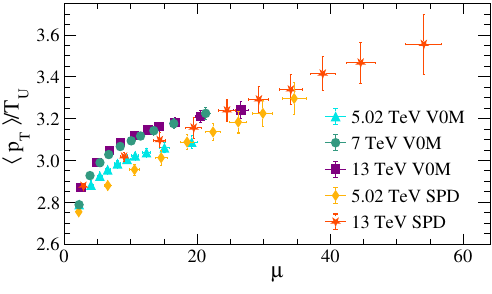}
    \caption{ $\avg{p_T}/T_U$ as a function of $\mu$ for all the analyzed data sets. 
    }
    \label{fig:mupTT}
\end{figure}

We calculate the 
mean transverse momentum by setting $n=1$ in Eq.~\eqref{eq:momentslab2}, resulting in 
\begin{equation}
        \avg{p_T} 
     = \frac{(q-1)(3q-5)}{(2-q)(2q-3)} \left( \frac{\Gamma\left( \frac{1}{q-1}\right) }{\Gamma\left( \frac{1}{q-1} - \frac{1}{2}\right)} \right)^2 T_U.
    \label{eq:avgpTe}
    \end{equation}
Note that $\avg{p_T}$ is proportional to $T_U$, where the proportionality constant only depends on $q$.
In the thermal limit $q\to 1$, Eq.~\eqref{eq:avgpTe} reduces to $\avg{p_T} = 2T_U$, describing a purely thermal system.

In Fig.~\ref{fig:meanpT}, we show the results of the $\avg{p_T}$ as a function of the temperature. In all cases, the $\avg{p_T}$ increases as the temperature does. In particular, this increment is more pronounced for the SPD classifier than for the V0M's.
We also computed the $\avg{p_T} $ trend by considering the $q$ and $\sigma$ 
as a function of the temperature given by Eqs.~\eqref{eq:sT} and~\eqref{eq:qT}.

We observed that the highest values of $\avg{p_T}$ correspond to the systems with the largest 
multiplicities, for which we found the largest values of $q$.  
However, in all cases, we found that $q>1$, meaning a non-negligible contribution of the produced particles that come directly from rare events \cite{Herrera:2024tyq}. Those events are 
taken into account by assuming the 
nonextensive description of the string tension fluctuations, where strong interactions lead to large momentum transfer, resulting in the high $p_T$ and 
heavy particle production \cite{Herrera:2024tyq}. This is aligned with other descriptions assuming nonextensivity for describing the systems created in high multiplicity pp collisions \cite{Braun-Munzinger:2000csl,Biro:2020kve,Badshah:2023ffj}.

In Fig.~\ref{fig:mupTT}, we show the ratio $\avg{p_T}/T_U$ for all the experimental data analyzed, which deviates from $3$, the value estimated for the medium form in central heavy ion collision for systems undergoing to a very quick thermalization \cite{Schlichting:2019abc,Gardim:2019xjs}.

\subsection{Variance and kurtosis of the transverse momentum spectrum}

\begin{figure*}[ht!]
    \centering
\includegraphics{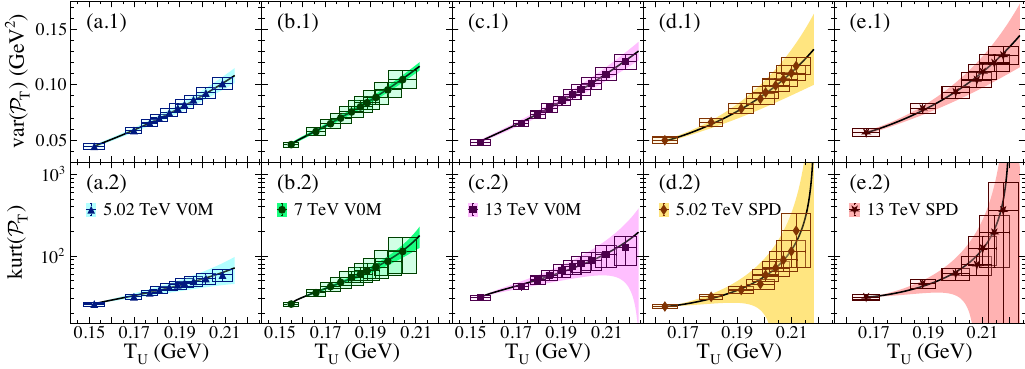}
\caption{Temperature dependence of variance (top panels)
, and kurtosis (bottom panels) 
of the transverse momentum spectrum of the charged particles produced in pp collisions at (a.1)-(a.2)  
    $\sqrt{s}=$ 5.02 TeV (V0M classes), (b.1)-(b.2)  $\sqrt{s}=$ 7 TeV (V0M classes), (c.1)-(c.2) $\sqrt{s}=$ 13 TeV (V0M classes), (d.1)-(d.2) $\sqrt{s}=$ 5.02 TeV (SPD classes), and (e.4)-(e.2) $\sqrt{s}=$ 13 TeV (SPD classes). Lines, figures, and colors are the same as in Fig. \ref{fig:meanpT}.}
    \label{fig:TMDstats}
\end{figure*}

The variance of the $p_T$ spectrum is computed in the standard form, 
var$(\mathcal{P}_T)=\langle \mathcal{P}_T^2 \rangle-\langle \mathcal{P}_T \rangle^2$, obtaining \cite{Herrera:2024zjy}
\begin{equation}
       \frac{ \text{var}(\mathcal{P}_T) }
       {T_U^2}
     =  \frac{2(q-1)}{3-2q} \left[ \frac{  \Gamma\left( \frac{1}{q-1} \right) }{ \Gamma\left( \frac{1}{q-1} - \frac{1}{2} \right) }\right]^2 - \left[\frac{4 - 2q}{5 - 3q} \right]^2 .
\label{eq:varpT}     
    \end{equation}
We point out that the variance over the square of the temperature is a function of $q$. Interestingly, this ratio tends to 1 in the thermal limit \cite{Herrera:2024zjy}.

It is important to extend the analysis of the $p_T$ spectrum beyond the variance estimation and consider the information from the heavy tail to study the $p_T$ spectrum shape. In particular, the kurtosis measures the contribution of the tails to the total probability, which corresponds to the observation of rare events that dominate the production of high $p_T$ particles. We compute the kurtosis as the fourth standardized moment, defined as
\begin{equation}
      \text{kurt}(\mathcal{P}_T) =  
    \frac{ \avg{\mathcal{P}_T}^4  
    \left[ \frac{\avg{\mathcal{P}_T^4}}{\avg{\mathcal{P}_T}^4} -   \frac{4\avg{\mathcal{P}_T^3}}{\avg{\mathcal{P}_T}^3} +  \frac{6\avg{\mathcal{P}_T^2}}{\avg{\mathcal{P}_T}^2} - 3\right]
   }{ \left[ \text{var}(\mathcal{P}_T) \right]^2 } . 
    \label{eq:kurt}
\end{equation}

Note that the calculation of $\avg{\mathcal{P}_T^4}$ from Eq.~\eqref{eq:ptn} requires $1<q<4/3$ for a finite estimation of the kurtosis 
when considering the integration range from 0 to infinity. 
This restriction is not presented in the kurtosis computation from the experimental $p_T$ spectrum because it is reported in a limited $p_T$ range.

We remark that in the thermal limit, the kurtosis tends to 9, meaning that the high $p_T$ particle production rate is not affected by increments in the center of mass energy or the multiplicity of the system.  

In Fig.~\ref{fig:TMDstats}, we show the behavior of the $p_T$ spectrum variance (top panels) and kurtosis (bottom panels) as a function of the temperature for both the V0M and SPD classifiers at $\sqrt{s} =$ 5.02, 7 (only V0M), and 13 TeV. 
In Fig.~\ref{fig:TMDstats} (bottom panels), we present the $\text{kurt}(\mathcal{P}_T)$ as a function of the temperature for all the analyzed datasets. We also consider the parametrization of $q$ and $\sigma$ in terms of the temperature (see Eqs.~\eqref{eq:sT} and~\eqref{eq:qT}) for estimating the trend of the variance and kurtosis.

Note that the highest multiplicity values are very close to the limit value at which kurtosis diverges. In consequence, there is a substantial increment in the uncertainty propagation in the data sets and the trend lines as shown in Fig.~\ref{fig:TMDstats} panels (c.2)-(e.2). 
In fact, for the I' SPD class at 5.02 TeV and the III', II', and I' SPD classes at 13 TeV, the $\avg{\mathcal{P}_T^4}$ is not finite.

In both cases, variance and kurtosis exhibit an increasing trend with the temperature, with a more pronounced concave-up behavior observed for the SPD classification. This indicates that the $p_T$ spectrum becomes flattened at low $p_T$ with an enhancement of the probability of producing high $p_T$ particles as the temperature or the multiplicity increases.

\section{Dependence of the Shannon entropy and heat capacity on the multiplicity classifiers}
\label{sec.HC}

In 1948, C. E. Shannon introduced the quantity $\mathcal{H}=-\langle \ln P(x)  \rangle$ as a measure of the information content in a probability distribution describing the random variable $x$ \cite{shannon1948mathematical}. 
In Ref.~\cite{Herrera:2024zjy}, the authors studied the Shannon entropy of the normalized $p_T$ spectrum for the charged particle production in minimum bias pp collisions. 
The study of the Shannon entropy can provide information regarding the contribution of the soft and hard parts.
In particular, for the thermal distribution, the Shannon entropy is $1+\ln T$ \cite{Herrera:2024zjy}. 
However, we expect deviations from the latter since the experimental data of the $p_T$ spectrum exhibits a power law tail.
In those cases, the increment of the Shannon entropy not only comes from a flattening of the soft part but also increases with an enhancement of the high $p_T$ particle production, which may differ from one event classification to another as a consequence of the selection bias.

We computed the Shannon entropy by considering the normalized $p_T$ spectrum as follows 
\begin{equation}
\mathcal{H} = -\int_0^\infty 
\left(\frac{1}{I_0}\frac{dN}{dp_T^2}\right) \ln\left(\frac{1}{I_0}\frac{dN}{dp_T^2}\right)   d p_T,
\label{eq:Entropy}
\end{equation}
where
\begin{equation}
    I_0=\int_0^\infty \frac{dN}{dp_T^2}  dp_T 
    = \frac{\sigma}{(2-q) \Gamma(a)} \sqrt{\frac{ q-1}{2} }
\end{equation}
is the normalization constant of Eq.~\eqref{eq:TMDU}. 
This marks a natural way of computing Eq.~\eqref{eq:Entropy} by considering $p_T$ as the random variable, according to what is usually done in the generalized ensemble theory \cite{niven2010jaynes,langen2015experimental}.
Since the description of the $p_T$ spectrum through \eqref{eq:TMDU} involves the Tricomi's function, the computation of the Shannon entropy must be performed by using numerical methods.

\begin{figure}
    \centering
    \includegraphics{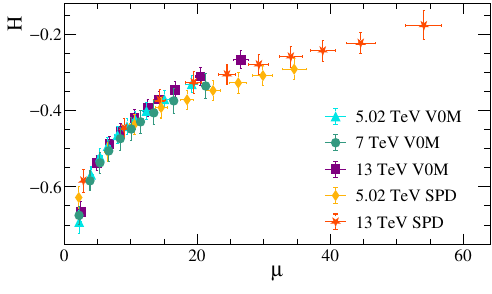}
    \caption{Shannon entropy as a function of the average multiplicity density for all the analyzed data sets.}
    \label{fig:Hmult}
\end{figure}

\begin{figure*}
       \centering
\includegraphics{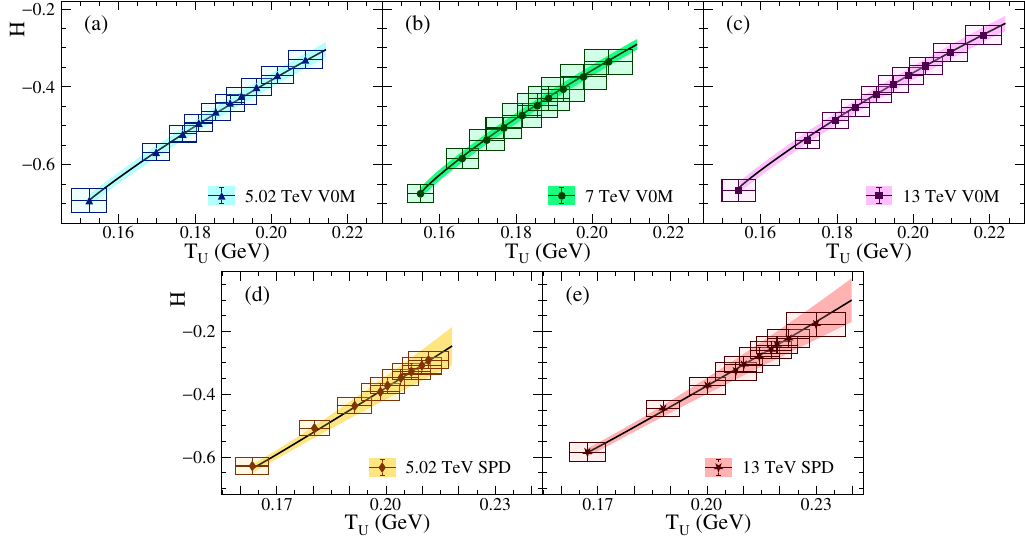}
\caption{Shannon entropy as a function of the temperature for: (a) $\sqrt{s}=$ 5.02 TeV (V0M), (b) $\sqrt{s}=$ 7 TeV (V0M), (c) $\sqrt{s}=$ 13 TeV (V0M), (d) $\sqrt{s}=$ 5.02 TeV (SPD), and (e) $\sqrt{s}=$ 13 TeV (SPD). The markers correspond to the computed values of the Shannon entropy using Eq.~\eqref{eq:Entropy} for each data set. In all cases, the solid lines correspond to the estimation of $\mathcal{H}$ considering the parametrizations of $\sigma$ and $q$ in \eqref{eq:sT} and \eqref{eq:qT}, respectively. The shaded regions correspond to the uncertainty propagation.}
\label{fig:H}
\end{figure*}

Figures~\ref{fig:Hmult} and~\ref{fig:H} show our estimations of the Shannon entropy for the $p_T$ spectrum data sets analyzed as a function of the average multiplicity density and the temperature, respectively. 
In all cases, $\mathcal{H}$ is a monotonically increasing function. In Fig.~\ref{fig:Hmult}, the Shannon entropy shows very similar behavior for the SPD and V0M classifiers. 
In contrast, Fig.~\ref{fig:H} shows a concave-down increasing behavior with the temperature for the V0M classes, which is not observed for the SPD classifier. 

Note that the increment in the V0M signal corresponds to an increment of the forward-backward activity, which 
has a nontrivial response in the shape of the $p_T$ spectrum measured in the midrapidity region \cite{ALICE:2013tla,Nijs:2023bzv}.
On the other hand, for the SPD classifier, the Shannon entropy increases with the multiplicity, which is expected because the SPD classifier is directly defined through the number of tracklets detected in the midrapidity region. Thus, the production of charged particles with high $p_T$ is enhanced as the multiplicity increases, raising the value of $q$ and, therefore, the temperature value.

We also analyze the rate of change of the Shannon entropy as a function of temperature to determine the subtle differences present in the multiplicity classifiers. This is done by computing the heat capacity, which is defined in thermodynamics as follows
\begin{equation}
    C = T_U\frac{d\mathcal{H}}{dT_U} ,
    \label{eq:heatdef}
\end{equation}
which is interpreted as a measure of how much heat is required to \emph{warm up} the system at which the particles are produced.
From the definition of $T_U$ in Eq.~\eqref{eq:TU}, the temperature grows with increments of $\sigma$ and $q$, but in all cases, it occurs simultaneously as a function of the average multiplicity. 
In this way, \emph{heating} the $p_T$ spectrum means a flattening of the soft part (production of soft particles in a larger low $p_T$ region) and an enhancement of the $p_T$ spectrum tail (higher rate of producing high $p_T$ particles).

\begin{figure*}
       \centering
    \includegraphics{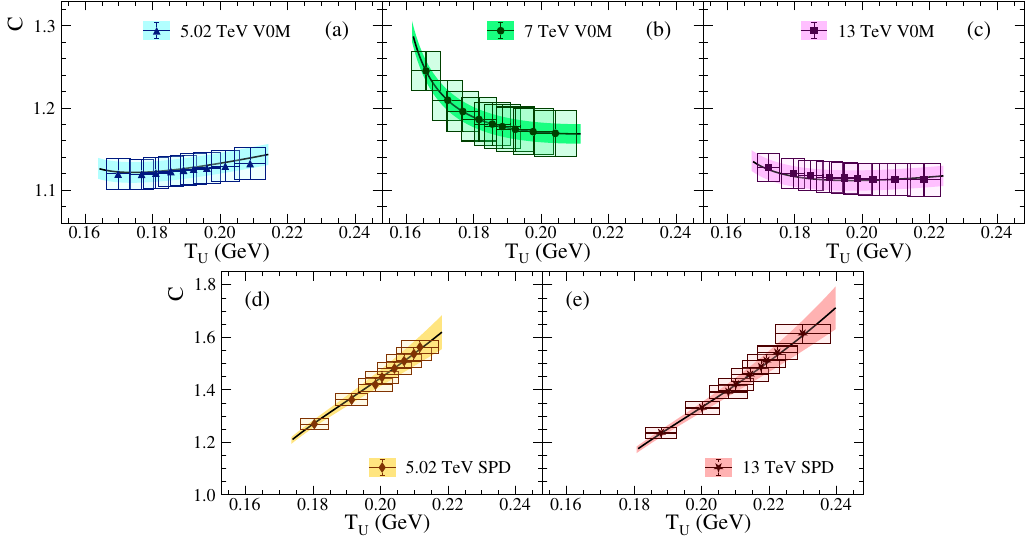}
\caption{Heat capacity as a function of the temperature for: (a) $\sqrt{s}=$ 5.02 TeV (V0M), (b) $\sqrt{s}=$ 7 TeV (V0M), (c) $\sqrt{s}=$ 13 TeV (V0M), (d) $\sqrt{s}=$ 5.02 TeV (SPD), and (e) $\sqrt{s}=$ 13 TeV (SPD). The markers correspond to the computed values of the heat capacity using the values of $q$ and $\sigma$ and the derivative with respect to the temperature of their parametrizations 
for each data set. In all cases, the solid lines correspond to the estimation of $C$ considering the only parametrizations of $q$ and $\sigma$. The shaded regions correspond to the uncertainty propagation.}
\label{fig:C}
\end{figure*}

The computation of the heat capacity \eqref{eq:heatdef} demands knowing the explicit or implicit dependence of the Shannon entropy on the temperature.
To do this, we use the explicit dependence of $\sigma$ and $q$ as a function of $T_U$, given by Eqs.~\eqref{eq:sT} and \eqref{eq:qT}, respectively.
In general, the heat capacity can not be expressed as a closed formula for the Tricomi's function \eqref{eq:TMDU}. In fact, it requires the computation of some derivatives of the $U$ function and the integration of terms of the form $U\ln U'$, as discussed in Ref.~\cite{Herrera:2024zjy}. In this way, we also numerically compute the heat capacity for all cases.
Figure~\ref{fig:C} contains our estimations of the heat capacity for the analyzed data.

We found that the heat capacity is sensitive to the multiplicity classifier.
The heat capacity for the SPD classification increases as the multiplicity does, as expected. It means the system requires more energetic collisions to produce more charged particles and, at the same time, with higher $p_T$ values.
Interestingly, the heat capacity for the V0M classes shows a decreasing concave-up behavior more pronounced for pp collisions at 7 TeV, which could be attributed to the more constrained pseudorapidity cut ($|\eta|<0.5$).
The main implication is that the system demands more energetic collisions to lower the system's temperature, which is opposite to the SPD classifier.
This occurs because the V0M classes are defined over distributions of tracklets that are not mutually exclusive.
Moreover, the increasing concave-behavior of the $q$ value for the V0M classes is a direct indication that the system produces higher $p_T$ particles with a significantly lower rate as the average multiplicity increases. 
This behavior impacts the heat capacity at higher temperatures and makes the system less accessible at lower-temperature states.

\section{Discussion}
\label{sec.Disc}

In this manuscript, we have used a nonextensive approach to study how the $p_T$ spectrum of charged particle production evolves with multiplicity in pp collisions. 
Where the data are categorized using two different multiplicity estimators defined by the ALICE Collaboration: V0M and SPD.
The model considers that the $p_T$ spectrum arises from the fragmentation of color strings with heavy-tailed tension fluctuations.
In particular, if the string tension fluctuations follow Eqs.~\eqref{eq:qGauss} and~\eqref{eq:PHag}, the $p_T$ spectrum becomes the Tricomi function~\eqref{eq:TMDU} and the Tsallis distribution~\eqref{eq:qexp}, respectively.
Interestingly, both distributions resemble an exponential decay at low $p_T$, allowing us to define the temperature as the inverse of the decay constant, as usual.
On the other hand, the $U$ and Tsallis distributions exhibit power-law tails at the high $p_T$ regime, modulated by $q$ and $q_e$, respectively.
Thus, the increase of the $q$ and $q_e$ values is directly linked to the enhancement of high $p_T$ particle production, which refers to a hardening of the $p_T$ spectrum.
Notably, the nonextensive approach accurately described the experimental data across the entire $p_T$-range, indicating that the majority of the particle production in pp collisions originates from string fragmentation. This is in agreement with the consensus that the collective phenomena 
have minimal influence on the $p_T$ spectrum in pp collisions~\cite{Nagle:2018nvi}.

We established the correlation between the temperatures and nonextensivity parameters of both descriptions as part of our analysis.
For instance, $q_e$ shows a linear correlation $q = 0.53 q_e + 0.46$, as seen in Fig.~\ref{fig:parcor} (a), although $q_e$ consistently remains below 1.2, within the expected range discussed in Ref.~\cite{Bhattacharyya:2017cdk}.
The soft scales $T_U$ and $T_e$ are proportional, with $T_e = 0.87 T_U$, as shown in Fig.~\ref{fig:parcor} (b). 
This is consistent with previous findings showing that $P_e(x)$ has a smaller variance than the $q$-Gaussian distribution at lower tensions, explaining the underestimation of $T_e$ relative to $T_U$~\cite{Herrera:2024tyq}. 
Given the correlations discussed above, it was sufficient to carry out the analysis using only the description based on the Tricomi’s function.

We observed that, in all the analyzed cases, $q$ increases monotonically with the temperature $T_U$, as shown in Fig.~\ref{fig:qT}.
The latter implies that high $p_T$ particle production is not suppressed, which is consistent with the lack of conclusive evidence of the jet quenching phenomena in pp collisions~\cite{Nagle:2018nvi,ALICE:2023plt}.
This behavior contrasts with that observed in AA collisions, where $q$ decreases with multiplicity, aligned with the suppression of high $p_T$ particles in larger systems~\cite{Pajares:2022uts,Garcia:2022eqg}.

We found significant differences in the dependence of $q$ as a function of $T_U$ across the multiplicity classifiers.
For V0M classes, $q$ saturates, indicating a poor enhancement of the hard particle production with increments of the temperature or multiplicity.
On the other hand, $q$ is a concave-up function of $T_U$, meaning the hardening of the $p_T$ spectrum.
This contrast is supported by the definitions of the multiplicity estimators.
The classification using the mid rapidity signal (SPD classes) allows for harder spectra at high multiplicities because the \textit{tracklets} classification can distinguish very high multiplicity events that make classes with high $\avg{p_T}$.
For instance, the I'-V' classes of SPD of pp collisions at $\sqrt{s}=13$ TeV contain 1.419\% of the total events, while the I class of V0M corresponds to 1\% of the events, as the ALICE collaboration reported in Ref. \cite{ALICE:2019dfi}. 
However, the V0M class I (highest multiplicity) does not necessarily contain information on all the highest multiplicity events in the transverse region. This is because the V0M classes are constituted by tracklet distributions. Consequently, the mean $p_T$ takes smaller values due to the large contribution of low multiplicity events.
In contrast, the SPD classification discriminates events with higher multiplicity and, therefore, harder $p_T$ spectra.

We further studied the differences between the V0M and SPD classifiers by analyzing the Shannon entropy as a function of the multiplicity, which was computed by considering the normalized $p_T$ spectrum through the $U$ function (see Eq.~\eqref{eq:Entropy}).
Interestingly, $\mathcal{H}$ increases monotonically with multiplicity. 
This behavior indicates that the $p_T$ spectrum becomes broader as a consequence of an increment of the temperature (flattening of the $p_T$ spectrum) and the raising of the probability of producing high $p_T$ particles (hardening of the $p_T$ spectrum).
However, the growth of the entropy as a function of temperature differs depending on the event classifier. 
Similar to $q$, $\mathcal{H}$ shows signs of saturation for the V0M classes, while for the SPD classes it follows a concave-up increasing behavior.

Due to the different responses of the Shannon entropy as a function of the multiplicity in the different classifiers, we investigated the rate of change of entropy with temperature through the heat capacity, defined in Eq.~\eqref{eq:heatdef}. 
This methodology allowed us to characterize in more detail the deviations present in the evolution of the $p_T$ spectrum for the V0M and SPD multiplicity classifiers.
In the case of the V0M classes, there are temperature intervals where the heat capacity decreases (see top panels of Fig.~\ref{fig:C}).
In particular, for $\sqrt{s}=7$ TeV, the heat capacity is a monotonically decreasing function of $T_U$. 
An explanation for this phenomenon may come from the perspective of an asymmetric relaxation of the system's evolution produced by the event selection biases.
In contrast, for the SPD classes, the heat capacity always increases with temperature (see bottom panels of Fig.~\ref{fig:C}).
This is expected because the increments in the number of tracklets in the midrapidity region are directly associated with the high $p_T$ particle production.

The behavior of the heat capacity can be interpreted in the usual thermodynamic sense:
An increasing heat capacity indicates that more energy is required to \emph{heat up} the $p_T$ spectrum, as seen for the SPD classes. 
In contrast, a decreasing heat capacity reflects that higher-temperature configurations become more accessible than lower-temperature ones, as we obtained for the V0M classes. 

\section{Summary and outlook}
\label{sec.Conc}

In this work, we studied the dependence on the multiplicity of the average $p_T$, variance, kurtosis, Shannon entropy, and heat capacity of the $p_T$ spectrum of charged particles produced in pp collisions. 
To this end, we analyzed the data sets for the center of mass energies of 5.02, 7, and 13 TeV classified under the V0M and SPD multiplicity estimators reported by the ALICE Collaboration (see Table~\ref{tab:dsets} for detailed information on the kinematic cuts of the analyzed data sets).
We fitted the Tricomi's function \eqref{eq:TMDU} and the Tasllis distribution \eqref{eq:qexp} to the experimental data of the $p_T$ spectra. 
We found that the nonextensivity parameters and temperatures of both descriptions are correlated. In particular, we studied the dependence of $q$ and $\sigma$ on the average multiplicity and the temperature $T_U$ (given by Eq.~\eqref{eq:TU}). 
We found that the behavior of $q$ is classifier-dependent: it saturates with V0M but it increases monotonically with SPD without saturation signals. 
The latter implies a stronger enhancement of hard particle production for the SPD classes.

In the computation of the $\avg{p_T}$, the variance, and the kurtosis of the $p_T$ spectrum considering the range from 0 to infinity, we found that all quantities are increasing functions of the temperature $T_U$ and multiplicity, reflecting the increase of the production of particles with higher $p_T$.
In general, the increasing values of the statistics are consequence of the flattening of the soft region along with a lifting of the distribution's tail.
All of the above result in an increase in the information content in the entire $p_T$ spectrum as the classes reach higher multiplicities, which was evaluated by computing the Shannon entropy through the normalized $p_T$ spectrum described by the Tricomi's function. 
In all cases, $\mathcal{H}$ increases with the multiplicity, aligned with a hardening of the $p_T$ spectrum. However, for SPD classes, the Shannon entropy shows an increasing concave-up behavior. 
Meanwhile, for the V0M classes, the entropy shows an increasing concave-down behavior. 

We also estimated the heat capacity as the derivative of the entropy with respect to the temperature (see Eq.~\eqref{eq:heatdef}).
We found different behaviors of the heat capacity as a function of the multiplicity for the V0M and SPD classifiers. 
In the case of the V0M classes, our analysis revealed that it is easier for the system to get higher temperatures than to cool it down. 
Meanwhile, for the SPD classes, more energy is required to increase the temperature of the system. 
We must point out that the multiplicity classifier chosen to select events may induce different responses in estimating theoretical or phenomenological observables that could lead to misleading interpretations or conclusions.

Finally, an immediate extension of this work is the analysis of the $p_T$ spectrum of the changed particle production under different event shape selectors, such as the transverse spherocity, relative transverse activity, or flatenicity, among others.

\begin{acknowledgments}
This work was funded by Consejo Nacional de Humanidades, Ciencias y Tecnologías (CONAHCYT-México) under the project CF-2019/2042,
graduated fellowship grant number 1140160, and postdoctoral fellowship grant numbers 645654 and 289198.
\end{acknowledgments}

\bibliography{ref}

\end{document}